# Influence of the Surface of a Nanocrystal on its Electronic and Phononic Properties

*Nuri Yazdani[1], Deniz Bozyigit[1], Kantawong Vuttivorakulchai[2], Mathieu Luisier[2], Vanessa Wood[1*]*

[1] Labratory for Nanoelectronics, Department of Information Technology and Electrical Engineering, ETH Zurich, CH-8092 Switzerland, [2] Nano TCAD Group, Department of Information Technology and Electrical Engineering, ETH Zurich, CH-8092 Switzerland

**Abstract**

Over the past thirty years, it has been consistently observed that surface engineering of colloidal nanocrystals (NC) is key to their performance parameters. In the case of lead chalcogenide NCs, for example, replacing thiols with halide anion surface termination has been shown to increase power conversion efficiency in NC-based solar cells. To gain insight into the origins of these improvements, we perform *ab initio* molecular dynamics (AIMD) on experimentally-relevant sized lead sulfide (PbS) NCs constructed with thiol or Cl, Br, and I anion surfaces. The surface of both the thiol- and halide-terminated NCs exhibit low and high-energy phonon modes with large thermal displacements not present in bulk PbS; however, halide anion surface termination reduces the overlap of the electronic wavefunctions with these vibration modes. These findings suggest that electron-phonon interactions will be reduced in the halide terminated NCs, a conclusion that is supported by analyzing the time-dependent evolution of the electronic energies and wavefunctions extracted from the AIMD. This work explains why electron-phonon interactions are crucial to charge carrier dynamics in NCs and how surface engineering can be applied to systematically control their electronic and phononic properties. Furthermore, we propose that the computationally efficient approach of gauging electron-phonon interaction implemented here can be used to guide the design of application-specific surface terminations for arbitrary nanomaterials.

The efficiency of semiconductor nanocrystal (NC) devices has benefitted tremendously from advances in NC surface chemistry.[1,2] High photoluminescent quantum yields in NCs are linked to engineering of the NC surface, where core/shell structures have been used[3] to optimize radiative rates,[4] emission wavelengths,[5] and hot carrier cooling rates.[6,7] NC-based solar cell performance has been improved with the evolution of termination strategies to increase charge carrier mobility,[8,9] reduce charge carrier recombination,[10] and increase carrier extraction through optimizing inter-NC band alignment.[11]

To date, the benefits resulting from surface engineering have predominately been discussed in terms of modification of the electronic properties. The introduction of a shell on a NC can be used to confine the electron, hole, or exciton in the core of the NC, an engineering approach to increase radiative recombination rates or decrease non-radiative rates by for example spatially separating the electron and hole from electronic defects stemming from surface states,[6,12] decreasing the rates of nonradiative Auger processes,[13] or minimize exciton polarization in the presence of an electric field.[14] In the case of NC solar cells, most literature emphasizes the role of surface termination on the passivation of defect states at the surface of NCs.[10,15,16]

Focus in the NC community is now shifting from descriptions of the static electronic properties of NCs to mechanisms based on the dynamic electronic and vibrational properties. Recent experimental[13,17-23] and theoretical work[24-27] points to efficient intraband carrier cooling and carrier trapping in NCs, mediated by phonons. In a recent study, we combined Inelastic Neutron Scattering (INS), ab-initio Molecular Dynamics (AIMD), and Thermal Admittance Spectroscopy (TAS) to show that low energy, large displacement phonon modes originating from the NC surface couple strongly to the electronic states and enable efficient multi-phonon-mediated non-radiative transitions.[17] In this context, here we aim to demonstrate how surface termination affects the vibrational properties of the NC and their influence on the electronic dynamics that play a role in device performance.

A typical approach to estimate the rates of multiphonon transitions involves the dimensionless Huang-Rhys factor, SHR, which indicates the strength of electron-phonon interactions to first order in atomic displacements. Transitions involving p phonons will have rates $k \propto (S_{HR})^p$. SHR is in general given by

$$S_{HR} = A \times M \times \langle u_\omega^2 \rangle \times (k_B T)^{-1} \times (\hbar \omega)^{-1} \times D^2, \qquad (1)$$

where $A$ is the unit-less overlap integral of the electronic states with $M$ number of phonon modes with frequency $\omega$, $D$ the deformation potential in (eV/Å), and $<u_\omega^2>$ is the mean square thermal displacement of the modes (Å$^2$).[28]

For experimentally-relevant sized NCs, the direct approach of computing $S_{HR}$ to the accuracy achieved via density functional theory is impractical. In the absence of inversion symmetry with respect to each of the $N$ atoms in the NC, calculation of the linear deformation potential, $D$, would require $6N$ energy calculations. In addition, given our previous findings[17] of large thermal displacement of the atoms in the NCs, a linear approximation to the deformation potential may not be sufficient for computation of multiphonon mediated transition rates $k$.

We therefore turn to ab-initio molecular dynamics (AIMD), previously employed for small CdSe clusters.[27,29] AIMD provides rapid insight of the phononic properties of NCs and the influence of the vibrational structure on the electronic wavefunctions. AIMD outputs could be used to perform stochastic surface hopping simulations to approximate carrier cooling dynamics, as performed on PbSe and CdSe NCs,[24,25] but to date such calculations have been limited to small clusters (< 150 atoms). Here, we show that one can gauge the strength of electron-phonon coupling and its impact on phonon mediated electronic transition rates through the investigation of the thermal fluctuation in state energies and the temporal dephasing of the band edge wavefunctions.

In this work, we specifically explore how surface termination selection influences phonon mediated processes in lead sulfide (PbS) NCs, which are used in LEDs, solar cells, photodetectors, transistors, and thermoelectrics.[30] We find that replacing thiol with halide ligands alters the electronic and phononic properties of the individual PbS NCs, reducing the overlap of the electronic wavefunctions with the large displacement surface phonon modes, decreasing electron-phonon coupling. We calculate thermal broadening of the electronic bandgap and gauge electronic transition probabilities by the extent of wavefunction dephasing, both of which indicate that phonon-mediated processes can be reduced via halide termination of NCs.

NCs are constructed according to the atomistic model introduced by Zherebetskyy *et al.*[31] As described in the Methods, we build NCs with a radius $r=\sim1.4$nm consisting of $N_{Pb}=201$ lead atoms and $N_S=140$ sulfur atoms, and terminated with $N_{Lig.}=120$ with methane-thiol (mth), Iodide (I), Bromide (Br) or Chloride (Cl) ligands (see **Figure 1a**). Mth is chosen instead of other thiols such as ethane-dithiol or 3-mercaptopropionic acid, which are used for NC devices,[32-34] since mth should maintain the same chemistry in its bonding to the NC surface yet has fewer number of atoms and therefore improves calculation speed. In addition, mth is monodentate in contrast to dithiols, which can bind mono or bidentate to the surface of the NCs and thereby increases the complexity of the assumptions required.[35] Geometry optimization, ground state electronic structure calculations, and AIMD in the canonical ensemble are performed on the various NCs within the CP2K program suite.[36] Details of the simulations are given in the Methods.

We first consider the effect of surface termination on the phonon density of states ($g(\omega)$). We extract the $g(\omega)$ from AIMD simulations as the power spectrum of the mass weighted position correlation function,[37]

$$g_i(\omega) = m_i \omega^2 \left| \mathcal{F}\{p_i(t)\} \right|^2, \quad g(\omega) = \sum_i g_i(\omega), \qquad (2)$$

where $m_i$ is the mass of atom $i$, $p_i(t)$ is the time trace of the position of atom $i$, and $\mathcal{F}$ represents a fourier transform. Partial density of phonon states can be investigated by summation over subsets of the atoms in the NC. We also calculate the mean square displacement of atom $i$, $\langle u_i^2 \rangle$. We previously showed that the $g(\omega)$ and $\langle u_i^2 \rangle$ extracted for bulk and nanocrystalline PbS with this approach is in good agreement with the values measured via experimental measurements of the $g(\omega)$ and $\langle u_i^2 \rangle$ using inelastic neutron scattering.[17]

As a result of the large mass disparity between the Pb and S atoms, in bulk PbS, the transverse acoustic (TA) and longitudinal acoustic (LA) modes, with peaks at 48.8cm$^{-1}$ and 84.4cm$^{-1}$ respectively, come almost entirely from the Pb atoms. The transverse optical (TO) and longitudinal optical (LO) modes originate predominantly from the S atoms and consist of sets of broad peaks at higher frequencies in the range of 100-200 cm$^{-1}$ (**Figure 1b**). The phonon bandstructure is provided in the **Supporting Information**.

For the NCs, regardless of surface termination, low frequency modes emerge below the TA peak of PbS bulk (48.8 cm$^{-1}$) and the LA peak shifts to lower energies (50 to 80 cm$^{-1}$). For the PbS/mth NC, the partial $g(\omega)$ for S-atoms extends to lower frequencies and the carbon (C) and hydrogen (H) atoms of the mth show vibrational modes in the range of 0 to 150 cm$^{-1}$. Due to the larger mass of Br and I compared to Cl, the partial $g(\omega)$ from the Br or I atoms is located at lower frequencies (0 to 150 cm$^{-1}$) while that for Cl extends over the entire range (0 to 250 cm$^{-1}$).

To understand the atomic origins of the $g(\omega)$ of the PbS NCs, we examine the partial $g(\omega)$ and $<u^2>$ for atoms located in three different regions in the NC (**Figure 2a**). The partial $g(\omega)$ for Pb and S atoms in PbS/mth and PbS/Cl NCs are shown in **Figure 2b**. As expected,[17] in the core of the NCs (Region 3), the partial $g(\omega)$ resembles that of bulk PbS while deviations from the bulk PbS $g(\omega)$ arise due to the outer atomic layers of the NC (Region 1). In Region 1, S atoms at the surface contribute in the low frequency region and Pb atoms at the surface contribute to high frequency region. This effect is particularly pronounced in the PbS/mth NC, while it is weaker in the Cl-terminated NC.

**Figure 2c** shows the $<u^2>$ at 100K for Pb, S, and halide anion atoms in the three NC regions are plotted for all four NC surface terminations. Values for Pb and S calculated for bulk PbS as provided for reference (black lines). Three trends are evident. First, the $<u^2>$ values for both Pb and S atoms decrease towards the value for bulk PbS the further the atoms are from the surface. Second, the $<u^2>$ values for the Pb and S atoms at the surface are smaller for PbS NCs with halide terminations than for the PbS/mth NC. Third, extremely large $<u^2>$ are observed for anions (S, I, Br, and Cl) on the [111] surface facets.

We can explain these observations by recalling that atomic displacement is inversely related to the effective spring constant of the phonon mode, $\kappa_{eff}$,[17]

$$\langle u^2 \rangle = k_B T / \kappa_{eff}. \tag{3}$$

$\kappa_{eff}$ is associated with the bonding strength between neighboring atoms. Pb and S atoms in the interior of a NC have a coordination number of 6, while the surface Pb and S atoms on the [100] facets of the NCs have a coordination number of 5, and surface anions such as S, I, Br, and Cl on the [111] facets have a coordination numbers of 2 or 3 (ignoring the S-C bonds in the PbS/mth NC). Thus larger $<u^2>$ are linked to decreased coordination number. Furthermore, the large $<u^2>$ of the Pb atoms on the outermost [111] facet for PbS/mth NC compared to halide terminated NCs can be explained by the strength of the bond. The effective spring constants associated with the Pb-X (X=mth, I, Br, Cl) bonds with the [111] surface atoms should scale with the percentage ionic character of the bonds,[38] and the decrease in $<u^2>$ with the increase in ligand electronegativity is therefore expected (**Table 1**).

Next we consider the electronic properties of the PbS NCs as a function of surface termination. In the bulk, PbS is a direct-gap semiconductor with the band gap occurring at the L-point, such that both the valence and conduction band are 4 fold-degenerate excluding spin (See **SI Fig S2**). The valence band maximum (VBM) states are of mixed 3*p*-S / 6*s*-Pb character, while the conduction band minimum (CBM) states are of mixed 6*p*-Pb / 3*s*-S character. For NCs, shape has been shown to effect the splitting and energetic ordering of quantum confined levels due to inter-valley coupling.[39-42] For our NCs, the 4-fold degeneracy of the VBM and CBM in bulk-PbS is broken due to the [100] facets of the NCs. Both the VBM and the CBM bands give rise to 3 degenerate states and a singly degenerate level, all with S-type envelope functions for the electron densities (**Figure 3a**). The CBM is singly degenerate (which we label $\psi_{1e}$) followed by triply degenerate states ($\psi_{2e\alpha}$), while the VBM is triply degenerate ($\psi_{1h\alpha}$) followed by a singly degenerate state ($\psi_{2h}$). The plot of number of electronic states (**Figure 3a**) indicates that the NCs exhibit no electronic states in the bandgap, which is to be expected for a defect free NC, and a larger density of states in the VB than in the CB, which is consistent with previous calculations on PbSe NCs[41] where this effect was attributed to stronger intraband coupling in the VB. In agreement with recent calculations by Voznyy *et al*.,[43] while the envelope function for all 8 band edge states are S-type, the underlying lattice wavefunctions have differing symmetries, and the lowest energy $\psi_{1h\alpha} \rightarrow \psi_{1e}$ transition has negligible optical coupling. Further discussion is given in the **Supporting Information**.

For our calculations, spin-orbit interactions (SO) and hybrid functional (HF) contributions are not included. To gauge how these effects may impact the electronic fine structure of the NCs, we compute the bulk bandstructure of PbS including SO interactions, and HF contributions (**Supporting Information**), and conclude that while both HF and SO must be included to achieve accurate bandgaps, the lowest energy CB and highest energy VB wavefunctions in the NC should not be significantly affected with SO+HF.

In **Figure 1c**, we plot cross-sections of the electron density for the VBM and CBM for NCs with the different surface terminations. The density plotted for the VBM is the average of the three-fold degenerate states, while for the CBM it is the density in the lowest singly degenerate state. Consistent with previous calculations on Cl-terminated NCs,[44] the wavefunctions for our halide-terminated NCs are confined in the NC away from the [111] facets. The total carrier density on the two outermost atomic layers on the [111] facet decreases by 40% (0.17→0.10) in the CBM and 60% (0.27→0.11) in the VBM from mth to Cl surface termination. The extent of electronic confinement increases with increasing electronegativity of the anion (S, I, Br, Cl), which is evidenced by the increasing electronic bandgaps (**Table 1**).

Combining our results for the electronic and phononic properties of the NCs indicates that halide anion surface termination should reduce electron-phonon coupling as the electronic wavefunctions are confined further away from the [111] surfaces, reducing the overlap of the electronic wavefunctions with the large thermal displacement surface phonons, and reducing the large mean thermal displacement of the Pb and S atoms close to the [111] surfaces. This provides an indication of how electron-phonon coupling changes as a function of surface termination. We define a parameter $A_{ue}$, the overlap of the carrier density with the mean thermal displacement of the atoms,

$$A_{ue} = \sum_i \left[ \langle u_i^2 \rangle \sum_\sigma |\varphi_{n,i\sigma}|^2 \right], \quad (4)$$

where $i$ runs over all atoms in the NC, and $\varphi_{n,i\sigma}$ are the components of the $n^{th}$ wavefunction projected onto the atomic orbitals $\sigma$ of the $i^{th}$ atom. The results at 100K for the CBM and VBM (averaged over the three degenerate VBM states) are tabulated in **Table 1.** This reduction in overlap indicates that the electron-phonon coupling should decrease as the effective electronegativity of the ligands increases.

We next assess how the reduction in electron-phonon coupling from halide surface termination of NCs affects phonon-mediated processes in the NCs such as thermal broadening of optical transitions due to fluctuations in energy and electronic transitions driven by phonon emission or absorption. To do so, we work with the time-dependent electronic wavefunctions and their energies sampled at each time step of the AIMD. As Supporting Information, we provide animations of the time dependent CBM wavefunctions and of the electron density in the 3 degenerate VBM states for each NC at the various temperatures.

As a first example, we consider the thermal broadening of time dependent optical band gap,

$$E_g(t) = E_{1e}(t) - E_{1h\alpha}(t). \tag{5}$$

The distributions of $E_g(t)$ for the PbS/mth and PbS/Cl NCs at 10K, 100K, and 300K are shown in **Figure 4a**, and the standard deviations of the distributions are given in **Table 2**. For both NCs, the $E_g(t)$ distributions broaden and blue shift with increasing temperature. The large thermal broadening is a typical feature of colloidal NCs,[7,45-47] and it has already been suggested to result from strong electron-phonon coupling.[46-50] Taking into account that the broadening should increase with a decrease in NC size,[7] the thermal broadening at 300K we extract agrees well with that measured on large PbS NCs.[46]

The red shift of the $E_g$ distributions with increasing temperature also stem from the electron-phonon interactions. Changes to $E_g$ with increasing temperature that arise from thermal expansion of the lattice are expected to lead to a blue shift for Pb-chalcogenide NCs.[47,51] We compute the electronic structure of the PbS/Cl NC with the atoms fixed at their mean positions calculated from the AIMD at 300K. The band gap increases from $E_g(0K)=1.29$ eV to $E_g(300K)=1.31$ eV, thus the observed red shift in the $E_g(t)$ distributions results from the lattice dynamics. While the red shift should be visible in temperature-dependent absorption of smaller

PbS NCs, it may not be in photoluminescence spectra because our calculations are performed for electrons in both the VBM and CBM. The energy fluctuations caused by Coulombic coupling to an electron in the VBM flip their sign when coupled to a hole.

As expected, **Figure 4a** shows that the broadening and blue shift, which are related to electron-phonon coupling, are significantly reduced in the PbS/Cl NC. This computational finding could explain the decrease of the thermal broadening upon halide passivation that has recently been demonstrated experimentally.[43]

As a second example, we examine the influence of surface termination on phonon-mediated electronic transitions. Here we focus on transitions to/from the VBM and CBM. For each, we calculate the wavefunction overlap autocorrelation function, $R_n(\tau)=|<\psi_n(t)|\psi_n(t-\tau)>|^2$, which indicates the rate and extent to which $\psi_n(t)$ dephases. In general, the rates for transitions to and from $\psi_n(t)$ will depend on $R(\tau)$. For example (within the single particle approximation implementation of the fewest switches surface hopping time domain density functional theory), the probability of transition to/from $\psi_n(t)$ from/to state $\psi_k(t)$ within a time step of $\Delta t$ is proportional to $(R(\Delta t))^{1/2}$.[25,52]

We thus use $R(\tau)$ to gauge how efficiently phonons can drive electronic transitions. For the CBM, the wavefunction overlap autocorrelation function is given as

$$R_{CBM}(\tau) = \left\langle \left|\langle\psi_{1e}(t)|\psi_{1e}(t-\tau)\rangle\right|^2 \right\rangle_t, \tag{6}$$

where the outer brackets, $<...>_t$, indicate averaging over the time trace of the AIMD. The VBM of the NCs are three-fold degenerate in the absence of atomic motion. During the AIMD, this degeneracy is lifted as the energies of the three states uniquely fluctuate, leading to avoided crossings in the energies of the 3 degenerate VBM states throughout the simulation and strong mixing of all three states with one another. For the VBM, we therefore calculate

$$R_{VBM}(\tau) = \frac{1}{3}\left\langle \sum_{\alpha,\beta} \left|\langle\psi_{1h\alpha}(t)|\psi_{1h\beta}(t-\tau)\rangle\right|^2 \right\rangle_t. \tag{7}$$

For bulk PbS, we average over the four fold degenerate CBM and VBM states:

$$R_{CBM,bulk}(\tau) = \frac{1}{4}\left\langle \sum_{i,j=1}^{4} \left|\langle \psi_{CBMi}(t)|\psi_{CBMj}(t-\tau)\rangle\right|^2 \right\rangle_t,$$

$$R_{VBM,bulk}(\tau) = \frac{1}{4}\left\langle \sum_{i,j=1}^{4} \left|\langle \psi_{VBMi}(t)|\psi_{VBMj}(t-\tau)\rangle\right|^2 \right\rangle_t. \quad (8)$$

In **Figure 4b**, we plot $R_{CBM}(\tau)$ and $R_{VBM}(\tau)$ for the PbS/mth and PbS/Cl NCs at 10K, 100K, and 300K. $R_{CBM}(\tau)$ and $R_{VBM}(\tau)$ for both NCs indicate strong initial dephasing, which levels out to mild fluctuations with a defined mean within the first ps. The initial dephasing time depends on what frequency phonon modes are coupled to the electronic wavefunctions. As we discuss below, the initial dephasing in the VBM stems from their larger coupling to the high-energy modes (150-250cm$^{-1}$). The extent of dephasing indicates how well phonons drive electronic transitions, with smaller mean values indicating a higher rate of transistion.

For both NCs, the extent of the dephasing of the VBM states is significantly larger than the CBM. This indicates faster phonon-mediated transition rates in the VB resulting from a denser DOS in the VB (**Figure 3a**), and the stronger coupling of the VBM states to phonons. This is consistent with recent experimental measurements, which indicate much faster carrier cooling rates in the VB compared to the CB in isotypic PbSe NCs.[53]

For all temperatures, the extent of the dephasing is significantly larger in the PbS/mth NC than the PbS/Cl NC. This indicates a significant reduction in the coupling of the phonons to electronic transitions upon Cl termination. In **Figure 4c**, we plot $R_{CBM}(\tau)$ and $R_{VBM}(\tau)$ for bulk PbS, and the PbSX (X=mth, Cl, Br, I) NCs at 100K. In both the CB and VB, the extent of the dephasing progressively increases from bulk to halide-terminated NCs to thiol-terminated NCs. In the CB, the distributions for the various halide anions overlap; however, in the VB, the PbS/I NC dephases to a greater extent relative to the PbS/Cl and PbS/Br. The extent of the dephasing is in good agreement with the ordering predicted by the overlap of the carrier density with the mean thermal displacement $A_{ue}$ (eq. (4), **Table 1**), again an indication that the dephasing is related to electron-phonon coupling in NCs.

To elucidate which phonons are responsible for the thermal broadening and dephasing of the electronic wavefunctions, we plot the spectral densities of $E_g(t)$ (**Figure 5a**) and $R_{CBM}(\tau)$ and $R_{VBM}(\tau)$ (**Figure 5b**), for the PbS/mth and PbS/Cl NCs at 100K. All three spectral density plots show coupling to the low (5-50cm$^{-1}$) and high (150-250cm$^{-1}$) energy modes, in agreement with our previous work,[17] where we experimentally showed that large energy multi-phonon-mediated electronic transition were driven by both low (25-50cm$^{-1}$) and high (120-240cm$^{-1}$) energy phonons. The spectral density of $E_g(t)$ reveals that reduction in thermal broadening for halide terminated NCs stems from a significant reduction in the coupling of $E_g$ to the high frequency optical modes compared to the PbS/mth NC. The spectral densities predominate contribution to the wavefunction dephasing comes from the low energy modes with smaller contributions from the high frequency modes, and a reduction in their coupling to the wavefunction dephasing with Cl termination is clear.

In conclusion, our work shows that electron-phonon interactions are reduced in halide terminated NCs compared to thiol-terminated NCs via (1) reduction of the mean square displacement of Pb and S atoms in the NC, and (2) confinement of the conduction and valence band wavefunctions away from the [111] surface where the largest atomic displacement modes occur. We further show that this reduction in electron-phonon interactions reduces thermal broadening of optical transitions and should suppress phonon-mediated electronic transitions. These findings explain the experimental results on thermal broadening, hot electron relaxation, and the improved performance of NC-based solar cells that used halide treatments and underscores the important and multifaceted role surface engineering can play optimizing NCs for different applications.

**Methods**

*Construction of NCs*

The approach used for the construction of the NCs investigated in this work is shown schematically in Figure S1. To construct the NCs, bulk rocksalt PbS (with a Pb or S atom centered on the origin) is cut along the eight (111) planes and six (100) planes at plane to origin distances (*r*) defined by the Wulff ratio $R_W$

$$r_{(1,0,0)} = AR_W, \quad r_{(1,1,1)} = AR_W^{-1}. \tag{9}$$

The scalar *A* is adjusted such that the resulting NC is S-terminated on the (111) facets. These (111)-surface terminating S atoms are then replaced with the desired ligand. To obtain an intrinsic semiconductor NC, overall charge balance must be maintained. This includes contributions from the ligands and charging of the NC compensated by counter-ions in solution [44,35],

$$\begin{aligned} N_{e(cat)} - N_{e(an)} &= 0, \\ 2N_{Pb} - 2N_S + N_L V_L + N_{ch} e &= 0, \end{aligned} \tag{10}$$

where $N_x$ refers to the number of Pb or S atoms, ligands (*L*), or surplus/deficiency of additional charges (*ch*), and $V_L$ is the valence of the ligand.

For a range of NC radii (~1nm to ~2nm), taking $R_W = 0.82$ and using thiol or halide anion ligands ($V_L = -1$), the charge balance condition (Equation 1.1) is almost satisfied ($\leq \pm 2e$). NCs cut with $R_W$ other than 0.82 require the removal of far more (typically >10) ligands/Pb-ligand-pairs or strong charging in order to satisfy eq. (10). For the case $R_W = 0.82$, to fully satisfy eq. (10), one to two ligands (-) or Pb-ligand-pairs (+) are removed. Alternatively, the NC can assume to be charged ($N_{ch} = \pm 1/\pm 2$). In contrast to the case of ligand removal, with chargin NCs retain octahedral symmetry. All of the NCs studied in this work are charged by +2e.

**Simulations**

Geometry optimization, electronic structure calculations, and *ab initio* molecular dynamics (AIMD) are performed within the CP2K program suite utilizing the quickstep module.[36] Calculations are carried out using a dual basis of localized Gaussians and plane waves,[54] with a 300Ry plane wave cutoff. As in previous calculations for CdSe[55] and PbS[17] NCs, Double-Zeta-Plus-Polarization (DZVP)[56], Goedecker–Teter–Hutter pseudopotentials[57] for core electrons, and the Perdew–Burke–Ernzerhof (PBE) exchange correlation functional are used for all calculations. Convergence to $10^{-8}$ in Self Consistent Field calculations is always enforced.

For NCs, non-periodic boundary conditions are used, and 4nmx4nmx4nm cubic unit cells are used for the $r$=1.4nm NCs. For bulk PbS calculations, periodic boundary conditions for the 4x4x4 cubic supercell (512 atoms) are used, and the supercell dimensions ($(6.0115)^3 Å^3$) are determined through a cell optimization using a conjugate gradient optimization. Geometry optimization is performed with the Quickstep module utilizing a Broyden–Fletcher–Goldfarb–Shannon (BFGS) optimizer. A maximum force of 24 meVÅ-1 is used as convergence criteria. All atoms in all systems were relaxed.

AIMD is performed in the canonical ensemble, using a CSVR thermostat, which achieves canonical sampling through velocity rescaling.[58] For thermalization and calibration of the thermostat, the time constant of the thermostat is set to 15 fs and the AIMD is run for 1ps. The time constant is then set to 1ps for the remainder of the AIMD. All AIMD steps prior to thermalization are discarded. AIMD time steps of 10fs are used for PbS/Cl, PbS/I, PbS/Br, and bulk, while 1.5fs time steps were employed for the PbS/mth NCs due to the high frequency C-H stretching modes on the mth ligands. We post-process the atom trajectories by removing the 6 macroscopic degrees of freedom (3x translation, 3x rotation) using the Iterative Closest Point algorithm of Besl and McKay.[59]

We follow the procedure described in our previous work[17] to compute the phonon band structure in Figure SI-1 and the density of states shown in Figure 1b. However, in this work we use a lattice constant value of 6.0115 Å, which is founds for bulk PbS via cell optimization in CP2K. More details of the bulk calculations are given in the **Supporting Information**.

**Wavefunction overlap calculations**

To save memory and computation time when calculating overlap integrals, the wavefunctions are rasterized into cubic ~(0.9Å)$^3$ voxels, which introduces errors into the overlap calculations. We estimate the magnitude of this error by computing inner products of the orthogonal atomic ground state wavefunctions $\left|\langle \psi_i^0 | \psi_j^0 \rangle\right|^2$. Due to the rasterization, the inner-products do not produce 0, but are consistently <10$^{-5}$, which we thus take as the intrinsic uncertainty in the overlap calculations.

ASSOCIATED CONTENT

**Supporting Information** The Supporting Information is available free of charge via the Internet at http://pubs.acs.org. Bulk PbS phonon bandstructure, electronic bandstructure with and without SO + HF, more analysis of the electronic fine structure of the PbS NCs, and figure describing NC construction, and captions for the GIF animations are provided.

**Animations** animations are available free of charge via the Internet at http://pubs.acs.org. Animations of the time dependent CBM wavefunctions and VBM carrier densities for each of the AIMD simulations is provided.

**Corresponding Author**

Email: vwood@ethz.ch

**Funding Sources**

The authors acknowledge ETH Research Grant (N.Y), Swiss National Science Foundation Quantum Sciences and Technology NCCR (D.B. and N.Y.), and the Swiss National Science Foundation project 149454/TORNAD (K.V.). Computations were supported by a grant from the Swiss National Supercomputing Centre (CSCS; project ID s674).

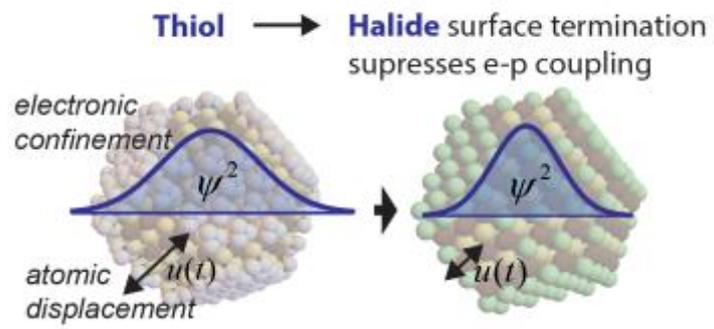

Table of contents figure

| $X$ | $\chi(X)$ | IC Pb-X (%) | $n_c[111]$ (e⁻) | | $<u_{[111]}>^2$ (Å²) | $A_{ue}$ (Å²e⁻) | |
|---|---|---|---|---|---|---|---|
| | | | CBM | VBM | | CBM | VBM |
| mth | 2.45 (ref.60) | 13.5 | 0.17 | 0.27 | 0.044 | 0.033 | 0.043 |
| I | 2.66 | 17.9 | 0.11 | 0.18 | 0.035 | 0.021 | 0.024 |
| Br | 2.96 | 23.9 | 0.10 | 0.14 | 0.037 | 0.019 | 0.021 |
| Cl | 3.16 | 27.6 | 0.10 | 0.11 | 0.035 | 0.019 | 0.020 |
| Bulk | - | - | - | - | - | 0.017 | 0.016 |

**Table 1.** Results for the electronic and phononic properties of the NCs with X-surface terminations. The effective electronegativity of the surface termination X in Pauling units ($\chi(X)$) and the percent ionic character of the X-Pb bond, computed by the Pauling formula $100(1-\exp[-1/4(\chi(X)- \chi(Pb))])$ are given for reference. The total carrier density ($n_c[111]$) on the outer Pb and X atoms on the [111] facet for both bands, and the mean square thermal displacement ($<u_{[111]}>^2$) at 100K of the same atoms are given. In the last column, the computed $A_{ue}$ values (eq. (4)) are shown for both bands at 100K.

|          | $\sigma_{Eg}$ (meV) | | |
|----------|-----|------|------|
|          | 10K | 100K | 300K |
| PbS/mth  | 13  | 31   | 72   |
| PbS/Cl   | 7   | 25   | 50   |

**Table 2.** Variance of the distribution of band gap energies, $E_{1e}(t)-E_{1h\alpha}(t)$, at various temperatures.

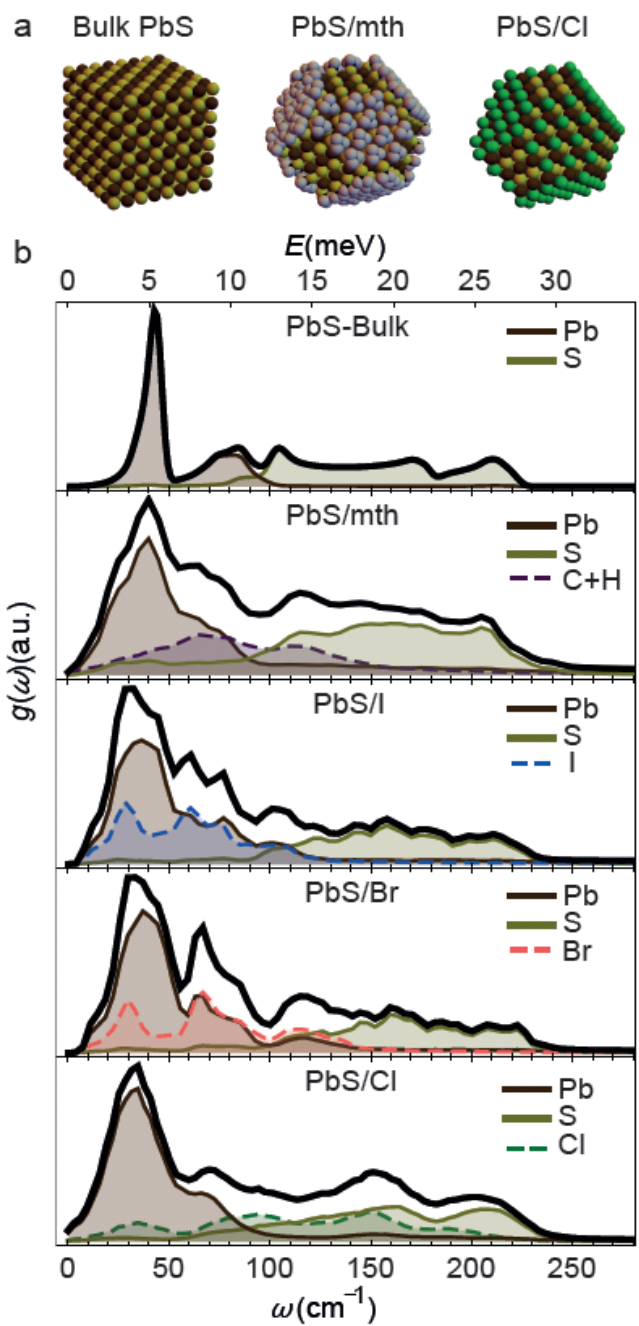

**Figure 1 Surface termination dependent phononic properties of PbS NCs.** a) Atomistic models of bulk PbS, a methane thiol-terminated PbS NC (PbS/mth), and a Cl-terminated PbS NC (PbS/Cl). b) Phonon density of states ($g(\omega)$) (black line) for bulk PbS (calculated in VASP) and for the mth-, I-, Br-, and Cl-terminated NCs (calculated via AIMD in CP2K). The partial $g(\omega)$ for Pb (brown shading), S (dark yellow shading), and the different terminations is also shown.

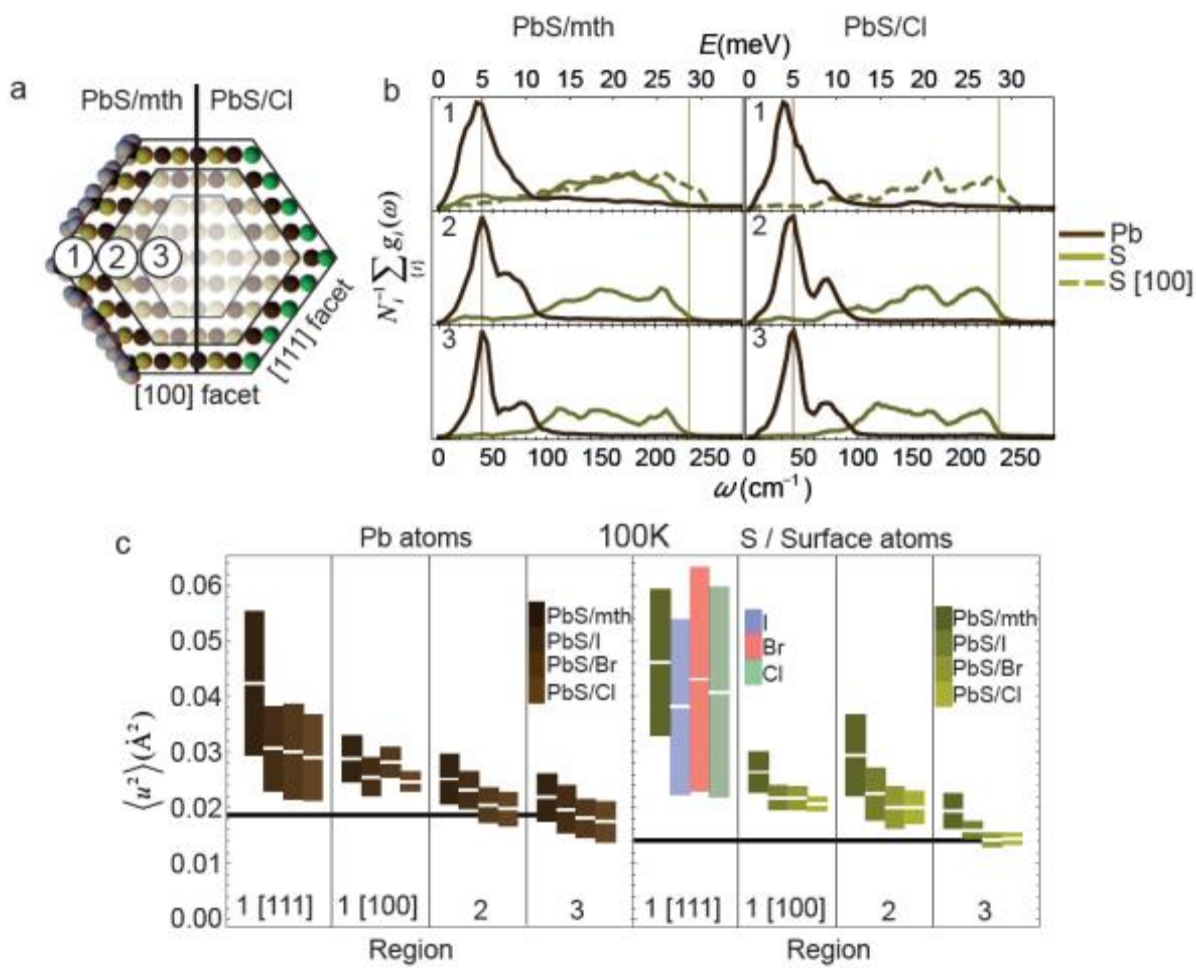

**Figure 2. Spatially resolved vibrational properties of atoms within the NC** a) Three regions of interest in the NC: the outer Pb and S layer (1), the sub-surface layers (2), and the core (3). The schematic emphasizes the different terminations on [100] and [111] NC facets. For all terminations, [100] facets are bare. For NCs with halide anions, there are no S atoms on the outer [111] surface. b) Partial $g(\omega)$ for Pb (brown line) and S atoms (yellow line) for each of the 3 regions of the PbS/mth (left) and PbS/Cl (right) NC. c) Plot of $<u^2>$ for Pb (left) and S, I, Br, and Cl atoms (right) in the three regions. [111] and [100] facets in Region 1 are plotted separately. The bars represent the range (-σ,+σ). The thick black lines are the $<u^2>$ calculated for Pb and S from bulk PbS (AIMD).

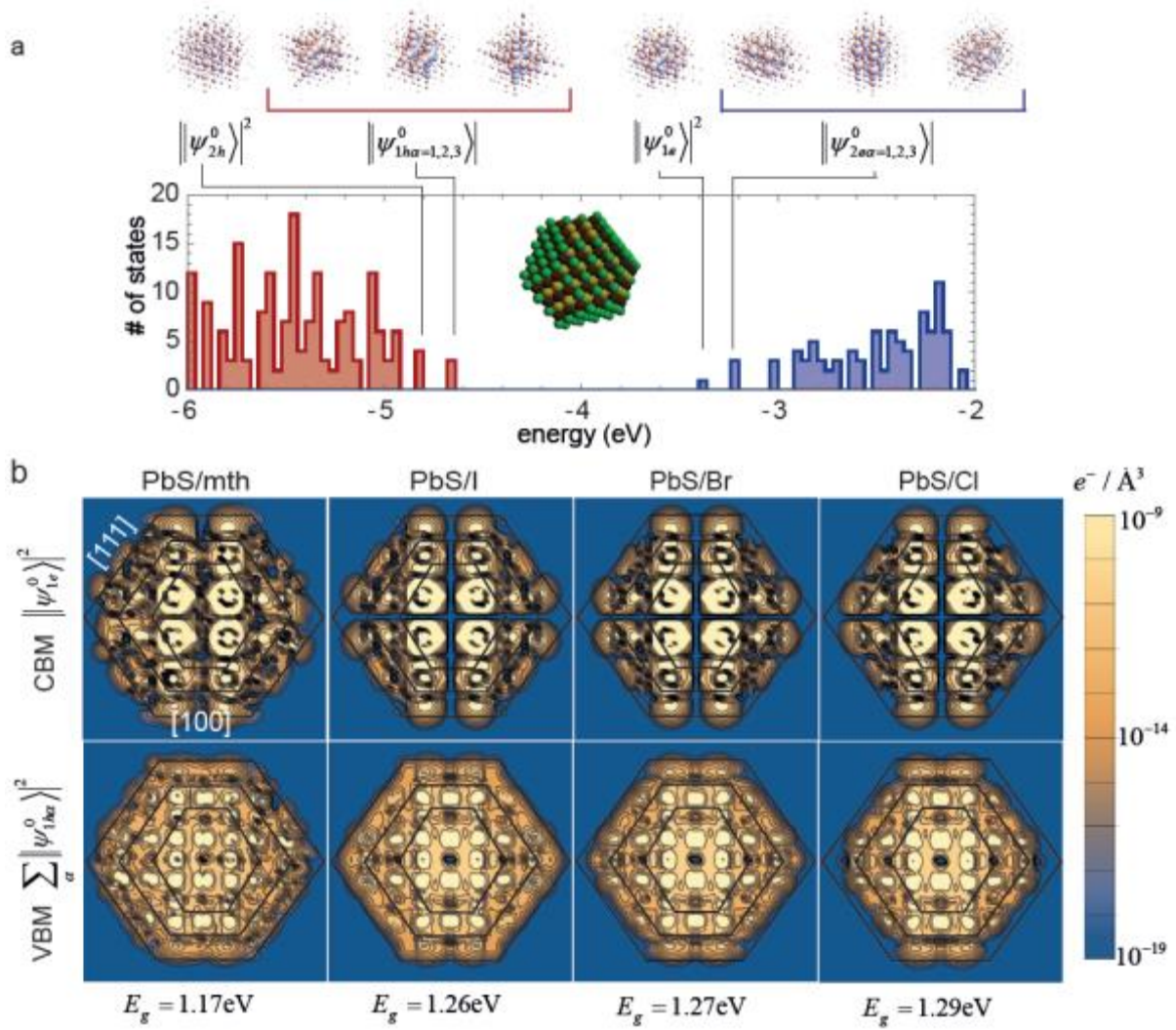

**Figure 3. Surface termination dependent electronic properties of PbS NCs** a) Number of electronic states in the valence band (VB) (red) and conduction band (CB) (blue) for a PbS/Cl NC plotted using 40meV binwidths. The electron densities of the first four valence states are shown above. b) Slices through a NC showing the electron density in the conduction band minimum (CBM) state and the 3-fold degenerate valence band maxima (VBM) states for all four surface terminations: mth, Cl, Br, and Cl. The total electron density per NC is 1. The calculated bandgaps of the NCs (indicated) are underestimated as is typical for density functional theory calculations.

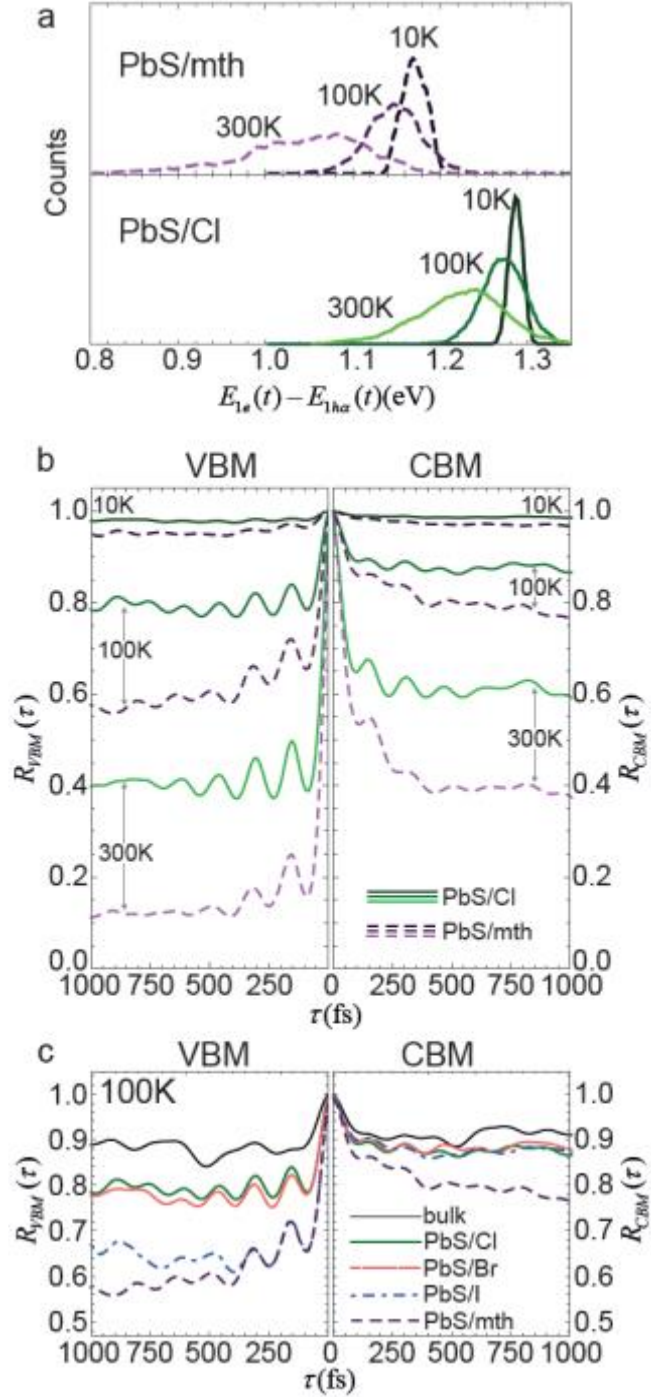

**Figure 4 Phonon induced thermal broadening and wavefunction dephasing** a) Histograms showing the band gap energy for a PbS/mth (top) and PbS/Cl (bottom) NC. b) Wavefunction overlap autocorrelation function for the CBM (eq. (6)) and VBM (eq. (7)) of the PbS/Cl and PbS/mth NCs at 10K, 100K, and 300K. c) Wavefunction overlap autocorrelation function for bulk PbS, halide terminated, and PbS/mth NCs at 100K (eqs. (6)-(8)).

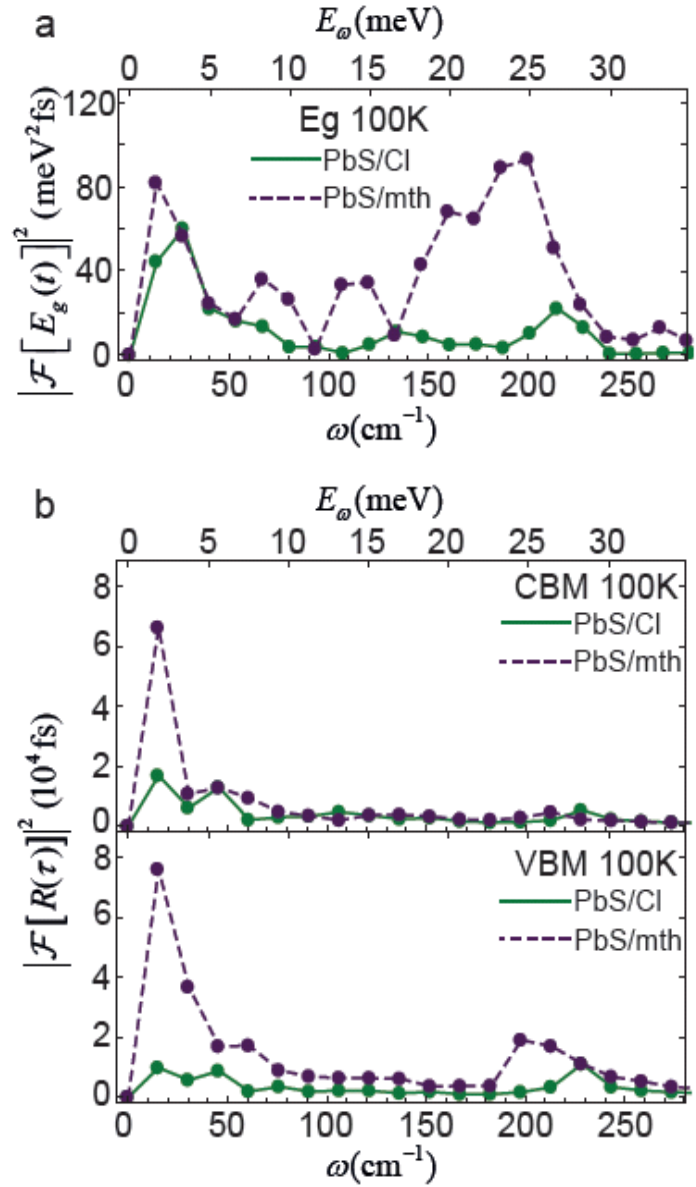

**Figure 5 Spectral Densities of thermal broadening and wavefunction dephasing** Spectral densities of (a) time dependent band gap energy ($E_g(t)$) and (b) the CBM and VBM wavefunction overlap autocorrelation functions for the PbS/mth (dashed purple lines) and PbS/Cl (solid green lines) NCs at 100K.